\newcommand{\bee}{\begin{equation}}
\newcommand{\ene}{\end{equation}}
\newcommand{\beea}{\begin{eqnarray}}
\newcommand{\enea}{\end{eqnarray}}

\documentclass[aps,preprint,pre,showpacs,superscriptaddress]{revtex4}
\usepackage{amsmath} %%%% To be able to use 'align' and 'subfigures', etc.
\usepackage{graphicx}% Include figure files
\usepackage{dcolumn}% Align table columns on decimal point
\usepackage{bm}% bold math
\begin{document}
\title{Quantized Plasmon Excitations of Electron Gas in Potential Well}
\author{M. Akbari-Moghanjoughi}
\affiliation{Faculty of Sciences, Department of Physics, Azarbaijan Shahid Madani University, 51745-406 Tabriz, Iran}

\begin{abstract}
Using the Schr\"{o}dinger-Poisson system in this paper the basic quantum features of plasmon excitations in a free noninteracting electron gas with arbitrary degeneracy is investigated. The standing wave solution of the free electron gas is derived from the corresponding linearized pseudoforce system with appropriate boundary conditions. It is shown that the plasmon excitation energies for electron gas confined in an infinite potential well are quantized eigenvalues of which are obtained. It is found that any arbitrary degenerate quantum electron gas possesses two different characteristic length scales, unlike the classical dilute electron gas, with the smaller length scale corresponding to the single particle oscillation and the larger one due to the collective Langmuir excitations. The probability density of the free electron gas in a box contains fine structures which are modulated over a larger pattern. The envelope probability density profile for the electron Fermi gas confined in an impenetrable well in different energy states are found to be quite similar to that of the free electron confined to an infinite potential well. However, the illustrative features of the plasmon theory presented in this research can be further elaborated in order to illuminate a wide range of interesting physical phenomenon involving both single particle as well as the collective features.
\end{abstract}

\date{\today}

\maketitle

\section{Historical Background}

Quantum behavior of atomic scale phenomena has fundamentally altered all our intuition of physical world, since the pioneering discoveries of Max Planck \cite{planck}, Erwin Schr\"{o}dinger \cite{es}, Louis de Broglie \cite{de} and many others. Almost all of the modern scientific developments in the fields like solid state \cite{kit,ash}, optoelectronics \cite{haug} and many technological achievements concerning miniaturized semiconductor integrated circuitry \cite{hu,seeg}, nano- and microelectronic device fabrication \cite{mark}, plasmonics \cite{man1}, etc. are indebted to our extensive knowledge of the quantum theory. Quantum effects appear when the characteristic length of the system compares to the thermal de Broglie wavelength of the particle leading to effective quantum interactions. Such interactions may be either due to the confinement of the particle in a potential well of length $l$ comparable to this wavelength or due to increase in the particle number density to the extreme limit where the interparticle distances compares to this quantum length so that the single particle wave functions in the system start to overlap \cite{bonitz}. However, there is fundamental difference between the later two cases as follows. The single particle Schr\"{o}dinger wave function possesses only one characteristic length determined by the width of the interacting potential well, whereas, the many particle wave function has two characteristic lengths with the smaller one corresponding to the single particle diffraction effect and the larger one due to the collective interactions. In plasmas the larger length scale is closely related to the well known Thomas-Fermi length. While the celebrated Schr\"{o}dinger equation satisfactorily deals with many natural phenomena regarding the noninteracting Bosonic particle systems, it may not be well suited for dense fermionic systems like quantum plasmas in which collective quantum electrostatic interactions are natural.

The field of quantum plasmas itself has a long development tradition. A big list of pioneering contributions to this field includes those of Fermi \cite{fermi}, Madelung \cite{madelung}, Hoyle and Fowler \cite{hoyle}, Chandrasekhar \cite{chandra}, Bohm \cite{bohm}, Pines \cite{pines}, Levine \cite{levine}, Klimontovich and Silin \cite{klimontovich} and many others which have led to discovery of many interesting collective quantum properties of dense and degenerate ionized environments. Quantum plasmas are relevant not only to metals, semiconductors and nanostructures but also are quite relevant to the astrophysical environments like big planetary cores, stellar media, white dwarfs and neutron star crusts. There are many recent literature which explore many interesting aspects of quantum plasmas \cite{man2,shuk1,manfredi,haas1,brod1,mark1,man3,asenjo,brod2,mark2,stern,ydj1,ydj2,dub1,shuk2,shuk3,shuk4,mark3,sarma,brod3,mold1,mold2,sm,akbrevisit}. In this paper we develop a method in order to solve for the quantized eigenvalues of the Schr\"{o}dinger-Poisson system for the electron gas of arbitrary degeneracy via the pseudoforce concept \cite{akbpseudo1}. We show that such a \textbf{plasmonic system} has two distinct characteristic lengths and when confined to the infinite potential well, similar to the quantum particle problem, leads to quantization of the plasmon excitation energies. Current results show akin similarities to the problem of particle confined to the infinite potential well and further reveals the fundamental features of collective effects in quantization of plasmon energy levels and properties of electron wave function in a free electron gas. The paper is organized as follows. The equation of state (EoS) of the finite temperature degenerate free electron gas is described in Sec. II. In Sec. III the hydrodynamic formulation is presented. The equivalent nonlinear Schr\"{o}dinger-Poisson (SP) formulation is given in Sec. IV. The coupled pseudoforce model \cite{akbpseudo2} and its standing wave solution is given in Sec. V. The numerical analysis and discussion is presented in Sec. VI and conclusions are drawn in Sec. VII.

\section{Equation of State of the Free Electron Gas}

Let us consider a free electron gas with an arbitrary degree of degeneracy characterized by the number density and statistical quantum pressure for electron fluid as follows
\begin{subequations}\label{np}
\begin{align}
&{n(\mu,T)} = \frac{{{2^{1/2}}m{^{3/2}}}}{{{\pi ^2}{\hbar ^3}}}  \int_{0}^{ + \infty } {\frac{{\sqrt{{\cal E}} d{\cal E}}}{{{e^{\beta ({\cal E}-\mu)}} + 1}}},\\
&{P(\mu,T)} = \frac{{{2^{3/2}} m{^{3/2}}}}{{3{\pi ^2}{\hbar ^3}}}\int_0^{ + \infty } {\frac{{{{\cal E}^{3/2}} d{\cal E}}}{{{e^{\beta ({\cal E} - {\mu})}} + 1}}.}
\end{align}
\end{subequations}
where $\beta=1/(k_B T)$ and ${\cal E}$ is the energy, $\mu$ is the chemical potential of electron gas and other parameters in (\ref{eos}) have their conventional physical meanings. The number density and pressure in (\ref{eos}) can also be rewritten in terms of familiar analytic $\text{Li}$ functions
\begin{equation}\label{pol}
{n(\mu,T)} =  - {\cal N}{\rm{L}}{{\rm{i}}_{3/2}}\left[ { - {\rm{exp}}\left( {-\beta {\mu}} \right)} \right],\hspace{3mm}{P(\mu,T)} =  - \frac{\cal N}{\beta}{\rm{L}}{{\rm{i}}_{5/2}}\left[ { - {\rm{exp}}\left( {-\beta {\mu}} \right)} \right],
\end{equation}
where ${\rm{Li}}_\nu(z)$ is the polylogarithm function of the order $\nu$ with the argument $z$. Also, the prefactor $\cal N$ appearing in (\ref{pol}) is defined as below
\begin{equation}\label{N}
{\cal N} = \frac{2}{{\Lambda^3}} = 2\left( {\frac{{{m k_B T}}}{{{2\pi\hbar^2}}}} \right)^{3/2},
\end{equation}
with $\Lambda$ being the electron de Broglie thermal wavelength. The EoS of isothermal free electron gas with arbitrary degeneracy is then written as \cite{ae}
\begin{equation}\label{eos}
P(\mu,T)=\frac{n(\mu,T)}{\beta}\frac{{\rm{L}}{{\rm{i}}_{5/2}}[ - \exp ({-\beta\mu})]}{{\rm{L}}{{\rm{i}}_{3/2}}[ - \exp ({-\beta\mu})]}.
\end{equation}
Note that in the classical limit $\beta\mu\ll -1$ we have ${\rm Li}_{\nu}[-\exp(-\beta\mu)]\approx-\exp(-\beta\mu)$ and the classical isothermal EoS, $P=nk_B T$ in (\ref{eos}) is retained. In the opposite fully degenerate extreme limit ($\beta\mu\gg 1$), we have ${\lim _{T \to 0}}{\mu} = {\epsilon_{F}}$ in which $\epsilon_F$ is the corresponding Fermi temperature and the polylog function is approximated as $\lim_{x\rightarrow\infty} {\rm Li}_{\nu}(-e^x)=-x^\nu/\Gamma(\nu+1)$. Therefore, in the fully degenerate isothermal free electron gas we have $P=(2/5)n k_B T_F$ in which $T_F=\epsilon_F/k_B$ is the Fermi temperature of the (zero temperature $T\ll T_F$) electron gas. The Fermi energy is related to the electron number density via $\epsilon_F=\hbar^2k_F^2/2m$ with $k_F=(3\pi^2 n)^{1/3}$ being the Fermi wavenumber. The typical value of the Fermi energy for metals range from $1$ to $12$eV. Using (\ref{pol}) it is confirmed that the identity $\partial P/\partial\mu=n(\mu,T)$ generally holds for the isothermal electron gas.

\textbf{In what follows we consider different models which describe collective quantum aspects of electron gas via the hydrodynamic coupling of quantum Bohm and electrostatic potentials. Some clarification and distinction between current models and the Hartree many body approximation, which assumes a description based on a single Slater determinant, usually considered in solid state literature, is in order. The complete many body description of a degenerate electron gas must be an idealistic formulation based on the N-electron wave function, $\psi(x_1,x_2,...,x_N,t)$ with all interactions between electron taken into account. As it is quite obvious this is a formidable task requiring extensive numerical work. However, in the weak correlation regime the electron gas is considered as essentially noninteracting (except the strong short range Pauli exclusion interactions dominating the system) where the many electron wave function can be decoupled into many single electron wave functions with one Schr\"{o}dinger equation governing each and with the common electrostatic potential of the gas coupling them to each other and to the Poisson's equation. Therefore, the following hydrodynamic and Schr\"{o}dinger-Poisson models are based on the Dawson's multistrem model \cite{man2}. Some recent developments based on the Wigner-Poisson and quantum hydrodynamic models appear in Refs. \cite{hurst,haaas} (see also the references therein).}

\section{The Hydrodynamic Model}

The dynamic properties of an electron fluid can be explored via the hydrodynamic model consisting of continuity, momentum balance and Poisson's equations, as follows
\begin{subequations}\label{hd}
\begin{align}
&\frac{{\partial {n}}}{{\partial t}} + \nabla  \cdot \left( {{n}{{\bf{u}}}} \right) = 0,\\
&m\left[ {\frac{{\partial {\bf{u}}}}{{\partial t}} + ({\bf{u}} \cdot \nabla ){\bf{u}}} \right] = \nabla \left( {e\phi  - \mu + e\phi_{ext}} \right) + \frac{{{\hbar ^2}}}{{2m}}\nabla \left( {\frac{{\Delta \sqrt n }}{{\sqrt n }}} \right),\\
&\Delta \phi  = 4\pi e {n},
\end{align}
\end{subequations}
in which we have used the identity $\nabla P(\mu,T) = n(\mu,T)\nabla \mu$ as an isothermal electron fluid EoS, described above. Unlike the electrostatic potential $\phi$ which is due to the electron gas itself, the component $\phi_{ext}$ can be any externally applied potential in the environment. The second term in the right hand side of the momentum balance equation denotes the quantum Bohm potential characterizing the electron quantum diffraction effect. The model (\ref{hd}) consists of a closed equation set to be used for full description of dynamics and evolution of electron plasma oscillations in an unmagnetized arbitrary degenerate fermionic gas. Note that in the zero temperature, $T\ll T_F$, limit the Fermi energy replaces the chemical potential in (\ref{hd}). At finite temperatures, the temperature dependence of the chemical potential may be obtained using the following Sommerfeld expansion in the limit $T< T_F$
\begin{equation}\label{cp}
\mu (T) = {\epsilon_F}\left[ {1 - \frac{{{\pi ^2}}}{{12}}{{\left( {\frac{T}{{{T_F}}}} \right)}^2} - \frac{{{\pi ^4}}}{{80}}{{\left( {\frac{T}{{{T_F}}}} \right)}^4} +  \cdots } \right].
\end{equation}

\section{The Schr\"{o}dinger-Poisson Model}

Let us now consider the 1D analogue of (\ref{hd}) model in the absence of external electrostatic potential $\phi_{ext}=0$. Thus, using the well-known Madelung transformation $\psi(x,t)=R(x,t)\exp[iS(x,t)/\hbar]$ in which $R(x,t)=\sqrt{n(x,t)}$ and $u(x,t)=(1/m)\partial S(x,t)/\partial x$ the 1D hydrodynamic model (\ref{hd}) can be cast into the following effective Schr\"{o}dinger-Poisson model \cite{manfredi} with a full nonlinear contribution to the field-density perturbations
\begin{subequations}\label{sp}
\begin{align}
&i\hbar \frac{{\partial \psi }}{{\partial t}} =  - \frac{{{\hbar ^2}}}{{2m}}\frac{{\partial {\psi ^2}}}{{\partial {x^2}}} - e\phi\psi  + \mu(R,T)\psi,\\
&\frac{{\partial {\phi ^2}}}{{\partial {x^2}}} = 4\pi e{n_0}R^2.
\end{align}
\end{subequations}
Note however that the appearance of $n_0$ which is the equilibrium number density of arbitrary degenerate electron gas in Poisson's equation indicates the normalization of $R$ with $R_0$ in the model. Note also that (\ref{sp}) has a potential beyond the original Schr\"{o}dinger equation by making contribution to the electrostatic interactions and well describes the plasmon wave packet evolution as will be consequently revealed.

\section{Pseudoforce Model and Standing Wave Solution}

We are now interested in obtaining the small amplitude standing wave solution of the free electron arbitrary degenerate gas confined in an infinite potential well of length $l$. To this end, let us consider the solutions of type $\psi(x,t)\equiv R(x)\exp[iS(t)/\hbar]$ and $\phi\equiv \phi(x)$. This ensures stationary wave types with standing $u=0$ character inside the potential well, as desired. The model (\ref{sp}) is then decomposed into the following coupled pseudoforce model
\begin{subequations}\label{pf}
\begin{align}
&\frac{{{\hbar ^2}}}{{2m}}\frac{{d^2{R(x)}}}{{d{x^2}}} + e\phi(x) R(x) - \mu (R,T)R(x) =  - \epsilon R(x),\\
&i\hbar \frac{{dS(t)}}{{dt}} = \epsilon S(t),\\
&\frac{{d^2{\phi(x)}}}{{d{x^2}}} = 4\pi e{n_0}{R^2}(x),
\end{align}
\end{subequations}
in which $\epsilon$ is the energy eigenvalue of the coupled Schr\"{o}dinger-Poisson system. The equation (\ref{pf}b) immediately yields $S(t)=\exp(-i\epsilon t/\hbar)$. In order for obtaining the small amplitude solution to the system (\ref{pf}) let us study the linearized system in which $\mu(R,T)\equiv \mu_0(R_0,T)$. Then we have the simplified following coupled differential equation system to solve with appropriate initial values
\begin{subequations}\label{lin}
\begin{align}
&\frac{{d^2 {R}}}{{d{x^2}}} + \frac{{2m(\epsilon - \mu_0 )}}{{{\hbar ^2}}}R - \frac{{2me}}{{{\hbar ^2}}}\phi  = 0,\\
&\frac{{d^2 {\phi}}}{{d{x^2}}} - 8\pi e{n_0}R = 0.
\end{align}
\end{subequations}
The coupled pseudoforce system (\ref{lin}) has the following simple solutions with the tentative initial conditions $R(0)=R'(0)=\phi'(0)=0$ and $\phi(0)=\phi_0$
\begin{subequations}\label{sol}
\begin{align}
&R(x) = {A_R}\left[ {\cos \left( {{k_+}x} \right) - \cos \left( {{k_-}x} \right)} \right],\\
&\phi(x) = {A_\phi }\left[ {\cos \left( {{k_-}x} \right) - {{\left( {{k_-}/{k_+}} \right)}^2}\cos \left( {{k_+}x} \right)} \right],
\end{align}
\end{subequations}

where $k_+$ and $k_-$ are the characteristic high and low plasmon wavenumbers, respectively. These parameters together with the corresponding solution coefficients are given below

\begin{subequations}\label{ka}
\begin{align}
&{k_+} = \alpha{k_p}\sqrt {1 + \sqrt {1 - \frac{{4{\epsilon_p^2}}}{{\delta {\epsilon^2}}}} },\hspace{3mm}{k_-} = \alpha{k_p}\sqrt {1 - \sqrt {1 - \frac{{4{\epsilon_p^2}}}{{\delta {\epsilon^2}}}} },\hspace{3mm}\alpha = \sqrt{\frac{\delta \epsilon}{2 \epsilon_p}},\\
&{A_R} = \frac{{e{\phi _0}/\delta \epsilon}}{{\sqrt {1 - 4{\epsilon_p^2}/\delta {\epsilon^2}} }},\hspace{3mm}{A_\phi } = \frac{{({k_+}/k_0)^2{\phi _0}}}{{(\delta \epsilon/{\epsilon_p})\sqrt {1 - 4{\epsilon_p^2}/\delta {\epsilon^2}} }},
\end{align}
\end{subequations}
where $\delta \epsilon=\epsilon-\mu_0$, $k_p=\sqrt{2m\epsilon_p}/\hbar$ and $\epsilon_p=\hbar\omega_p$ is the plasmon energy quanta with $\omega_p=\sqrt{4\pi e^2 n_0/m}$ being the electron plasma frequency and $n_0$ the unperturbed electron gas number density. Note that we use $\mu_0$ for the unperturbed chemical potential of the electron gas.

We may now proceed with quantizing the energy eigenvalues by confining the free electron gas in the infinite potential well of width $l$. It is then required that the real function $R(x)$ representing the number density of the electron fluid to vanish at the boundaries of the well. However, we use a more general periodic boundary condition as $(k_+\pm k_-)l=2\pi n$ with an arbitrary quantum integer number $n$, not to be confused with the electron gas number density. We then obtain the following quantized energy eigenvalue $\epsilon$,
\begin{equation}\label{en}
{\epsilon_n} = {\mu _0} + 2{\epsilon_p}\left( {1 + \frac{{\lambda_p^2{n^2}}}{{{2l^2}}}} \right),\hspace{3mm}{\lambda_p} = \frac{{2\pi }}{{{k_p}}}.
\end{equation}
At zero temperature, $T\ll T_F$, we have the limiting value of $\mu_0=\epsilon_F$. The quantized values of the characteristic wavenumbers then read
\begin{subequations}\label{ks}
\begin{align}
&{k_+} = {k_p}\sqrt {1 + \frac{{{\lambda_p^2}{n^2}}}{2l^2}} \sqrt {1 + \sqrt {1 - \frac{1}{{{{\left( {1 + {\lambda_p^2}{n^2}/2l^2} \right)}^2}}}} },\\
&{k_-} = {k_p}\sqrt {1 + \frac{{{\lambda_p^2}{n^2}}}{2l^2}} \sqrt {1 - \sqrt {1 - \frac{1}{{{{\left( {1 + {\lambda_p^2}{n^2}/2l^2} \right)}^2}}}} },
\end{align}
\end{subequations}
It should be noted that for any given value of $n$, the high an low wavenumbers $k_+$ and $k_-$ are reciprocal of each other, i.e., $k_+k_-=1$. In the normalized scheme $l=l/\lambda_p$, $x=x/\lambda_p$, $k_{\pm}=k_{\pm}/k_p$, $\epsilon=\epsilon/\epsilon_p$ and $\mu_0=\mu_0/\epsilon_p$ and using the normalization $\int_0^1 {{R^2(x)}dx}  = 1$ to define the parameter $\phi_0$ in (\ref{sol}), one arrives at following simplified results for energy values of the free electron gas
\begin{equation}\label{norm}
\delta {\epsilon _n} = {\epsilon _n} - \mu_0  = 2 + \frac{{{n^2}}}{{{l^2}}},\hspace{3mm}{k_{\pm} } = \sqrt {1 + \frac{{{n^2}}}{{2{l^2}}}} \sqrt {1 \pm \sqrt {1 - \frac{1}{{{{\left( {1 + {n^2}/{2l^2}} \right)}^2}}}} }.
\end{equation}

The normalized solutions corresponding to (\ref{sol}) are then given as follows
\begin{subequations}\label{solnorm}
\begin{align}
&{R_n}(x) = \left[ {\cos \left( {{k_{+}}x} \right) - \cos \left( {{k_{-}}x} \right)} \right] = 2\sin \left( {\frac{{{k_{+}} - {k_{-}}}}{2}x} \right)\sin \left( {\frac{{{k_{+}} + {k_{-}}}}{2}x} \right),\\
&{\phi _n}(x) = \left[ {\left( {1 + {n^2}/2{l^2}} \right) + \sqrt {{{\left( {1 + {n^2}/2{l^2}} \right)}^2} - 1} } \right]\left[ {\cos \left( {{k_ - }x} \right) - {{\left( {\frac{{{k_ - }}}{{{k_ + }}}} \right)}^2}\cos \left( {{k_ + }x} \right)} \right].
\end{align}
\end{subequations}
It is clearly rmarked that solutions $\psi_n(x,t)=R_n(x)\exp(i\epsilon_n t/\hbar)$ have two characteristic length scales $2\pi/k_{\pm}$ corresponding to high frequency single-particle and low frequency collective excitations. It is also confirmed that the free electron energy dispersion in the Fermi gas follows the familiar relation $\delta\epsilon {_n} = {\hbar ^2}{k_n^2}/2m$ with $k_n^2=k_{-}^2+k_{+}^2$ being the total quantized momentum. We then have formulated the double length-scale theory of the arbitrary degenerate free electron gas confined in a potential well exactly analogous to the standard quantum theory of particle in a box.

Current theory may be easily extended to 3D case with the isotropic electrostatic potential of the form $\phi(x)$ related to the common electron number density by Poisson's equation for all three independent dimensions. By using the standard separation of variables procedure (before the lumbarization) with the assumption $\psi(x,y,z)=\psi_1(x)\psi_2(y)\psi_3(z)$ in ($\ref{pf}$), three similar independent coupled linear equations for each dimension can be obtained as follows
\begin{subequations}\label{lin3}
\begin{align}
&\frac{{{d^2}R_i({x_i})}}{{d{x_i}^2}} + \frac{{2m(\epsilon{_i} - \mu )}}{{{\hbar ^2}}}R_i({x_i}) - \frac{{2me}}{{{\hbar ^2}}}{\phi_i}({x_i}) = 0,\\
&\frac{{{d^2}{\phi _i}({x_i})}}{{d{x_i}^2}} - 8\pi e{n_0}R({x_i}) = 0,\hspace{3mm}(i = 1,2,3).
\end{align}
\end{subequations}
where $x_1=x$, $x_2=y$ and $x_3=z$. Therefore, for an electron gas confined in a cubic box of sides $l$ one obtains the eigenvalues given in (\ref{norm}) with new definitions $n^2=n_x^2+n_y^2+n_z^2$ and $\delta \epsilon=\sum\limits_i {\epsilon_i}=\sum\limits_i {{\hbar ^2}k_i^2/2m}$ with $k_i^2=k_{i+}^2+k_{i-}^2$. The normalized wavefunction $\psi(x,y,z)$ assuming $\phi_{01}=\phi_{02}=\phi_{02}=\phi_0$ and taking other initial values to be zero is given below
\begin{equation}\label{sol3norm}
\psi (x,y,z) = 8\prod\limits_i {\left[ {\sin \left( {\frac{{{k_{i + }} - {k_{i - }}}}{2}{x_i}} \right)\sin \left( {\frac{{{k_{i + }} + {k_{i - }}}}{2}{x_i}} \right)\exp \left( {\frac{{i{\epsilon _i}t}}{\hbar }} \right)} \right]}.
\end{equation}
It is remarked that in 2D and 3D cases some of the plasmon wavefuntions become degenerate with the same energy values.

\section{Numerical Analysis and Discussion}

Figure 1 shows the variations of the number density and plasmon parameters with the chemical potential $\mu_0$ of the arbitrary degenerate free electron Fermi gas. Figure 1(a) indicates two distinct classical ($\mu_0<0$) and quantum ($\mu_0>0$) regimes for the number density of electrons. The large increase in the electron temperature only significantly increases the number density of electrons for chemical potential values around $\mu_0=0$. The plasmon energy variation with the chemical potential is shown in Fig. 1(b) indicating typical values of few electron Volts for metals (this value for the Aluminium is $15$eV) and a sharp increase with increase in the electron number density. It is also observed that the increase in temperature of the electron gas leads to significant increase in the value of plasmon energy for regions of low chemical potential values. Variations in the plasmon wavenumber and characteristic length with chemical potential are depicted in Figs. 1(c) and 1(d). It is remarked that the plasmon wavelength decreases sharply for denser gases particularly for lower gas temperature values.

Figure 2 shows the discrete energy values of plasmon excitation for different values of the normalized width of the infinite potential well. The $\delta\epsilon=0$ corresponds to the energy eigenvalue of $\epsilon=\mu_0$. It is remarked that the plasmon spectrum always has an energy gap of value $2\epsilon_p$. Therefore, the quantized plasmon energy are given by $\epsilon=\mu_0+2\epsilon_p+2\epsilon_e n^2$ in which $\epsilon_e$ is the ground state energy of an electron in an infinite well of same length and $n$ is the plasmon quantum number. It is also clearly evident that the density of states (DoS) at an specified energy range decreases as the width of the potential well decreases.

Figure 3 depicts the variations in the wavenumber eigenvalues of the confined electron gas. It is remarked that, compared to the case of single electron in a potential well, the electron gas possesses two characteristic lengths \cite{akbpseudo2} corresponding to wavenumbers $k_{\pm}$ as indicated in Fig. 3. The higher value $k_+$ corresponds to the single electron oscillation while $k_-$ is due to the collective oscillations of electron gas. It is also revealed that in the classical limit ($l\to\infty$) the two wavenumber values merge into single value. It is also seen that decrease of the width of the potential well leads to increase/decrease of the positive/negative wavenumber branch value. This feature is shown to become more significant for higher quantum number values. It seen that the value of $k_-$ vanishes at very small vales of normalized length $l$.

Figure 4 shows the wavefunction, probability density, electrostatic potential and electric field profiles in the infinite potential well in the ground state $n=1$. Figure 4(a) differs fundamentally from the ground state wavefunction of a single electron due to the influence of collective electrostatic effects of free electron gas. The probability density of the corresponding state depicted in fig. 4(b) many definite places inside the potential well where th electron number vanishes. Note that the integral $\int {{R^2}(x)dx}$ gives the fractional number of electrons distributed within the spacial range $dx$. The profile in Fig. 4(b) confirms that in the ground state of the free electron gas electrons are mostly distributed in the middle of the potential well. The later feature is quite similar to the ground state probability of finding a particle confines inside the impenetrable well. The electrostatic potential and field profiles shown in Fig. 4(c) and 4(d) reveal that the electrostatic energy inside the well maximized in the middle of the bix where mos of the particles exist in the ground state.

In Fig. 5 we have shown the probability distribution of electrons in ground state of electron gas in infinite potentials with different widths. As it is clearly evident, the decrease in the width of the potential leads to decrease in the number of probability density maxima at which electrons are accumulated. Further decrease of the well sized eventually can lead to the ideal quantum case of single electron confinement with a profile of a single maxima similar to the case of a particle in a 1D box.

Figure 6 depicts the waqvefunction profile of the free electron Fermi gas confined in a one dimensional (1D) infinite potential well at different quantum state. The spiky profiles indicate oscillations in the electron number through the potential well. However, the $n=2$ and other higher energy states have radical difference with the ground state one in that there are electron gas voids at broader spacial ranges at higher energy levels. However, the envelop of the wavefunction profiles at higher quantized energy values has a close resemblance to that of a particle in a box. The time evolution of the wavefunction for the 1D free electron gas at different quantum states is depicted in Fig. 7. Figure 8, on the other hand, shows the colorful wavefunction patterns for the a 2D free electron degenerate gas at it different plasmon energy state. Figures 8(c) and 8(d) show two different wavefunctions with the same plasmon energy, so called the degenerate states.

\textbf{Current model in its present form fully accounts for the long range Coulomb collective effects in the context of single-electron wave function approximation considered also in Hartree-Fock method. The later method however also accounts for the electrostatic interactions via the self-consistent field in the mean field approximation as well as other electron gas correlations called the Fermi correlations which appear as electron exchange effect. It is however a possible extension for current model to include the exchange correlation effect as density functional form which appears in recent quantum plasma literature \cite{hurst}. It should be noted that one of the greatest advantages of current model compared to the more general nonlinear models is that it provides important analytic results in the linear limit and a linear combination form for the electrostatic potential and wave function solution are presented for the electron gas system with arbitrary degree of degeneracy.}

\section{Conclusion}

Using the effective Schr\"{o}dinger-Poisson system we have studied the quantum properties of the free electron Fermi gas with arbitrary degeneracy and have found that the energy of the free electron degenerate gas confined to an impenetrable potential well is quantized. An energy gap approximately equal to two plasmon energy exist in the quantized spectrum of the gas. The system of degenerate quantum electron gas is shown to have two different characteristic length scale corresponding to single- and multi-particle effects. Current theory of plasmon excitations can be extended to investigate the plasmon interactions and propagation in different environments.

\end{document}